\documentclass[review,3p,authoryear,times]{elsarticle}
\usepackage{color}
\usepackage{multirow,setspace,times,amssymb,amsmath,graphicx,color,subfigure,url}
\usepackage{dsfont,bm}
\usepackage{threeparttable}
\usepackage{rotating}
\usepackage{lscape,pdflscape}
\usepackage{dcolumn}
\usepackage{listings}
\usepackage{relsize}
\usepackage{booktabs}
\usepackage[colorlinks=true,linkcolor=blue,citecolor=blue,urlcolor=blue]{hyperref}
\usepackage[absolute,overlay]{textpos}
\usepackage{tikz}
\usepackage{caption}
\usepackage{pdflscape}
\usepackage{float}
\usepackage{afterpage}     
\usepackage{placeins}      
\usepackage{lipsum}        
\usepackage[capitalize]{cleveref} 
\usepackage{pdfpages}
\usepackage{overpic}
\journal{Elsevier}

\begin{document}
	
\begin{frontmatter}
		
\title{Risk spillovers between the BRICS and the U.S. staple grain futures markets}

\author[SUIBE]{Ying-Hui Shao}
\author[SILC]{Yan-Hong Yang }
\author[SB,RCE,SM]{Wei-Xing Zhou \corref{cor1}}
\ead{wxzhou@ecust.edu.cn}
\cortext[cor1]{Corresponding author. }

\address[SUIBE]{School of Statistics and Information, Shanghai University of International Business and Economics, Shanghai 201620, China}
\address[SILC]{SILC Business School, Shanghai University, Shanghai 201899, China}
\address[SB]{School of Business, East China University of Science and Technology, Shanghai 200237, China}
\address[RCE]{Research Center for Econophysics, East China University of Science and Technology, Shanghai 200237, China}
\address[SM]{School of Mathematics, East China University of Science and Technology, Shanghai 200237, China}

\begin{abstract}
    This study examines contemporaneous and lagged spillovers in BRICS staple grain futures markets and their linkages with U.S. markets. Results show that contemporaneous spillovers dominate, while net spillovers are driven by lagged connectedness. Systemic risk is lower in intra-BRICS markets than in those including the U.S., highlighting the U.S. grain market's significant influence. Brazilian and U.S. grains are key net spillover contributors, excluding U.S. rice, while South African staple grains act as major net receivers. Notably, spillovers between soybeans are the strongest. The study also reveals heterogeneous impacts of the Russia-Ukraine conflict and Black Sea Grain Initiative on grain futures.
\end{abstract}
		
\end{frontmatter}

\section{Introduction}
\label{S1:Introduction}
	
Staple grains like wheat, corn, soybeans, and rice play a critical role in global trade, food security, and financial markets. Their futures markets are essential for price discovery, risk management, and the stability of global food supply chains \citep{Li-Chavas-2023-AmJAgrEcon,Zhou-Dai-Duong-Dai-2024-JEconBehavOrgan}. BRICS is not only the five largest and most promising emerging markets but also a leading producer and consumer of staple grains, exerting significant global influence \citep{Khalfaoui-Hammoudeh-Rehman-2023-EmergMarkRev}. Generally, U.S. futures prices are a benchmark for global grain commodity pricing \citep{Li-Hayes-2017-JFuturesMark}. Thus, understanding how connectedness occurs in BRICS staple grain futures markets and their cross-market linkages with the U.S. is crucial, as these spillovers influence global price stability, trade policies, and economic shifts \citep{Jiang-Todorova-Roca-Su-2017-ApplEcon}.
	
Prior studies have highlighted the heterogeneity and complexity of spillover effects in grain futures markets \citep{Han-Liang-Tang-2013-QuantFinanc,Chen-Weng-2018-EmergMarkFinancTrade,Zivkov-Kuzman-Subic-2020-AgricEcon,Hu-Zhu-Zhang-Zeng-2024-PlosOne}.	While some studies focus only on spillovers between specific pairs of countries, such as between China and the U.S., others broaden their scope to include multiple markets. \cite{Li-Hayes-2017-JFuturesMark} demonstrate that the U.S. soybean futures market remains the most influential in the long term, leading price changes in Brazil and China. \cite{Zhu-Dai-Zhou-2024-JFuturesMark} show that CBOT corn, soybean, and wheat serve as primary risk transmitters to DCE corn and soybean, underscoring the directional nature of Sino-US grain futures market dynamics. However, \cite{Han-Liang-Tang-2013-QuantFinanc} reveal bidirectional information transfer between the Chinese and U.S. soybean futures markets, with China's soybean futures prices influencing price discovery in the U.S., challenging the traditional view of China’s market as a satellite to the U.S.
Notably, current geopolitical tensions, particularly those arising from the Russia-Ukraine conflict, have further intensified the complexity and uncertainty of international grain futures market spillovers, with recent evidence indicating that spillovers vary significantly across different market conditions, while U.S. grain futures continue to serve as the dominant global price benchmark \citep{Zhang-Sun-Shi-Ding-Zhao-2024-HumSocSciCommun}.

The financialization of agricultural commodities enhances liquidity and market efficiency but also increases contagion risk, particularly in the context of the current challenges posed by complex geopolitics, climate change, and frequent trade tensions \citep{Goldstein-Yang-2022-JFinanc,Just-Echaust-2022-EconLett,Guo-Li-Zhang-Ji-Zhao-2023-JCommodMark,Wang-Dong-Sun-Shi-Ji-2024-EconModel,Zhang-Sun-Shi-Ding-Zhao-2024-HumSocSciCommun}. To this end, we examine the internal dynamics within BRICS and their external linkages with the U.S., analyzing differences in risk spillovers of staple grain futures between these two systems while considering the impacts of the Russia-Ukraine conflict and the Black Sea Grain Initiative (BSGI) \citep{Behnassi-ElHaiba-2022-NatHumBehav,Wu-Ren-Wan-Liu-2023-FinancResLett,Ni-Cherif-Chen-2024-FinancResLett,Urak-Bilgic-Florkowski-Bozma-2024-BorsaIstanbRev}. 

This study makes three primary contributions. First, it utilizes data on the four major staple grains critical to global food security and examines multiple spillover dimensions, including overall, contemporaneous, and lagged spillovers. Second, it provides new empirical evidence on the internal connectivity of key emerging grain futures markets and their cross-market risk dependencies with mature markets. Third, the findings on evolving shock transmission pathways have significant implications for policymakers, traders, and investors navigating the complexities of the global grain trade.

The paper is organized as follows: Section~\ref{S2:Data:Methodology} outlines the data and methodology,  Section~\ref{S3:Empirical analysis} presents empirical results and robustness checks, and Section~\ref{S4:conclusion} concludes.

\section{Data and methodology}
\label{S2:Data:Methodology}
\subsection{Data}

This study examines the logarithmic returns derived from the futures prices of staple grains originating from BRICS countries and the U.S., covering the period from December 1, 2020, to August 11, 2023. The data were retrieved from Wind and Bloomberg databases. Due to the unavailability of Russian grain futures data, Black Sea region staple grain futures traded on CBOT are used as a proxy for Russia. For consistency, each commodity is denoted by combining the country's ISO-2 code with the first letter of the grain. Specifically, BR, IN, CN, ZA, and US represent Brazil, India, China, South Africa, and the United States, respectively, while the first letters s, c/m, w, and r correspond to staple grains: soybean, corn/maize, wheat, and rice. For instance, BRs represents Brazil's soybean futures. Since the Black Sea region lacks an ISO-2 code, BS is used to represent it. The futures data are sourced from Brazil’s BMF exchange, India’s NCDEX exchange, South Africa’s SAFEX exchange, China’s DCE exchange, and CBOT, encompassing a total of 18 staple grain futures. Recognizing the significant influence of U.S. grain futures on the global market, we define two systems: System 1 includes all 18 staple grain futures; System 2 focuses solely on intra-BRICS markets, comprising 10 grain futures.

\begin{figure}[htp]
	\centering
	\includegraphics[width=0.5\linewidth]{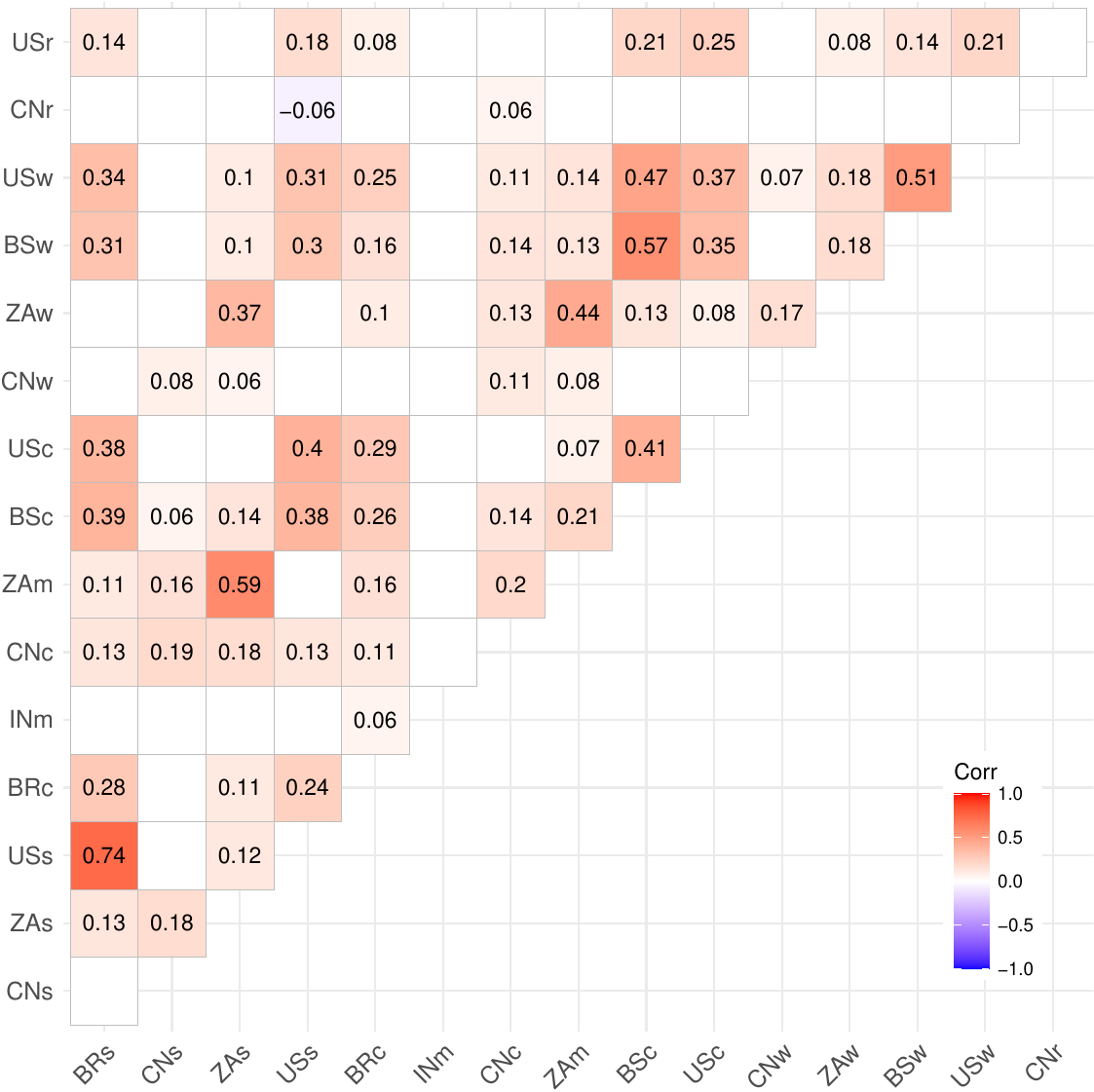}	
	\caption{Pairwise correlation heatmap. Notes: Coefficients that are not significant at the 10\% significance level are represented by blanks.}
	\label{Fig:Correlation:US}
\end{figure}

Table~\ref{Tb:Statistics} provides an overview of the descriptive statistics. With the exception of South African soybean futures, the returns of all grain futures do not pass the normality test. Additionally, all return series are found to be stationary. The vast majority of pairwise grain futures are positively correlated, peaking at 0.74 between Brazilian and U.S. soybeans, as shown in Fig.~\ref{Fig:Correlation:US}.

\begin{table}[htp]
	\centering
	\begin{threeparttable}
		\small
		\caption{Descriptive statistics of grain futures}
		\label{Tb:Statistics}
		\begin{tabular}{lllllll}
			\toprule
			Notation & Mean (x$10^3$) & SD & Skewness & Kurtosis & Jarque-Bera & ADF \\
			\midrule
			BRs & 0.178 & 0.014 & -1.134 & 12.038 & 2543.309*** & -9.175*** \\ 
			CNs & -0.163 & 0.011 & -0.053 & 5.989 & 262.040*** & -8.319*** \\ 
			ZAs & 0.666 & 0.013 & 0.054 & 3.146 & 0.960 & -8.042*** \\ 
			USs & 0.270 & 0.016 & -1.050 & 10.41 & 1737.264*** & -7.723*** \\ 
			BRc & -0.116 & 0.014 & 0.208 & 5.291 & 158.885*** & -7.829*** \\ 
			INm & 0.793 & 0.016 & 0.039 & 6.352 & 329.296*** & -7.601*** \\ 
			CNc & 0.016 & 0.008 & -0.394 & 4.465 & 81.056*** & -8.057*** \\ 
			ZAm & 0.466 & 0.016 & 0.139 & 3.333 & 5.506* & -9.180*** \\ 
			BSc & 0.110 & 0.012 & 0.868 & 9.94 & 1499.246*** & -7.818*** \\ 
			USc & 0.197 & 0.024 & -1.916 & 21.687 & 10658.812*** & -7.905*** \\ 
			CNw & 0.292 & 0.013 & 0.253 & 18.248 & 6817.827*** & -8.535*** \\ 
			ZAw & 0.368 & 0.008 & -0.007 & 4.858 & 101.142*** & -8.847*** \\ 
			BSw & 0.390 & 0.012 & 0.471 & 8.321 & 855.448*** & -6.613*** \\ 
			USw & 0.145 & 0.025 & -0.403 & 14.634 & 3983.961*** & -8.333*** \\ 
			CNr & -0.265 & 0.003 & 1.077 & 13.654 & 3460.656*** & -9.307*** \\ 
			USr & 0.351 & 0.014 & -1.462 & 19.344 & 8075.201*** & -9.923*** \\ 
			\bottomrule
		\end{tabular}
		\begin{tablenotes}
			\footnotesize
			\item Note: * , **, *** indicate significance at the 10\%, 5\%, and 1\% levels, respectively.
		\end{tablenotes}
	\end{threeparttable}
\end{table}

\subsection{Methodology}

This study employs a novel $R^2$ decomposition approach proposed by \cite{Naeem-Chatziantoniou-Gabauer-Karim-2024-IntRevFinancAnal} to analyze the overall, contemporaneous, and lagged spillover effects among the staple grain futures markets of the BRICS countries and the U.S. The total connectedness index (TCI) is defined as the average $R^2$ of the $k$ multivariate linear regressions:
\begin{equation}
	TCI=\frac{1}{K}\sum_{k=1}^{K}R_{k}^{2}=\left(\frac{1}{K}\sum_{k=1}^{K}\sum_{i=1}^{K}\bm{R}_{C,k,i}^{2,d}\right)+\left(\frac{1}{K}\sum_{k=1}^{K}\sum_{i=1}^{K}\bm{R}_{L,k,i}^{2,d}\right)=TCI^{C}+TCI^{L},
\end{equation}
where $ \bm{R}^{2,d}_C $ and $ \bm{R}^{2,d}_L $ represent contemporaneous and lagged spillovers, respectively.
The total directional spillovers to and from others, along with the net total spillover, are given by:
\begin{align}
	TO_{i} &= \sum_{k=1}^{K}\bm{R}_{C,k,i}^{2,d} + \sum_{k=1}^{K}\bm{R}_{L,k,i}^{2,d} = TO_i^C + TO_i^L \\
	FROM_{i} &= \sum_{k=1}^{K}\bm{R}_{C,i,k}^{2,d} + \sum_{k=1}^{K}\bm{R}_{L,i,k}^{2,d} = FROM_i^C + FROM_i^L \\
	NET_i &= TO_i - FROM_i \\
	&= (TO_i^C + TO_i^L) - (FROM_i^C + FROM_i^L) \\
	&= (TO_i^C - FROM_i^C) + (TO_i^L - FROM_i^L) \\
	&= NET^{C}_i + NET^{L}_i.
\end{align}
If $NET_i > 0$ ($NET_i < 0$), series $i$ is considered a net transmitter (receiver) of shocks \citep{Balli-Balli-Dang-Gabauer-2023-FinancResLett}. Likewise, net pairwise directional connectedness ($NPDC_{ij}$) functions similarly to $NET$, but at the pair level \citep{Cocca-Gabauer-Pomberger-2024-EnergyEcon}.

\section{Empirical analysis}
\label{S3:Empirical analysis}
	
\subsection{Average spillover effects}

\begin{figure}[htp]
    \centering
    \includegraphics[width=0.98\linewidth]{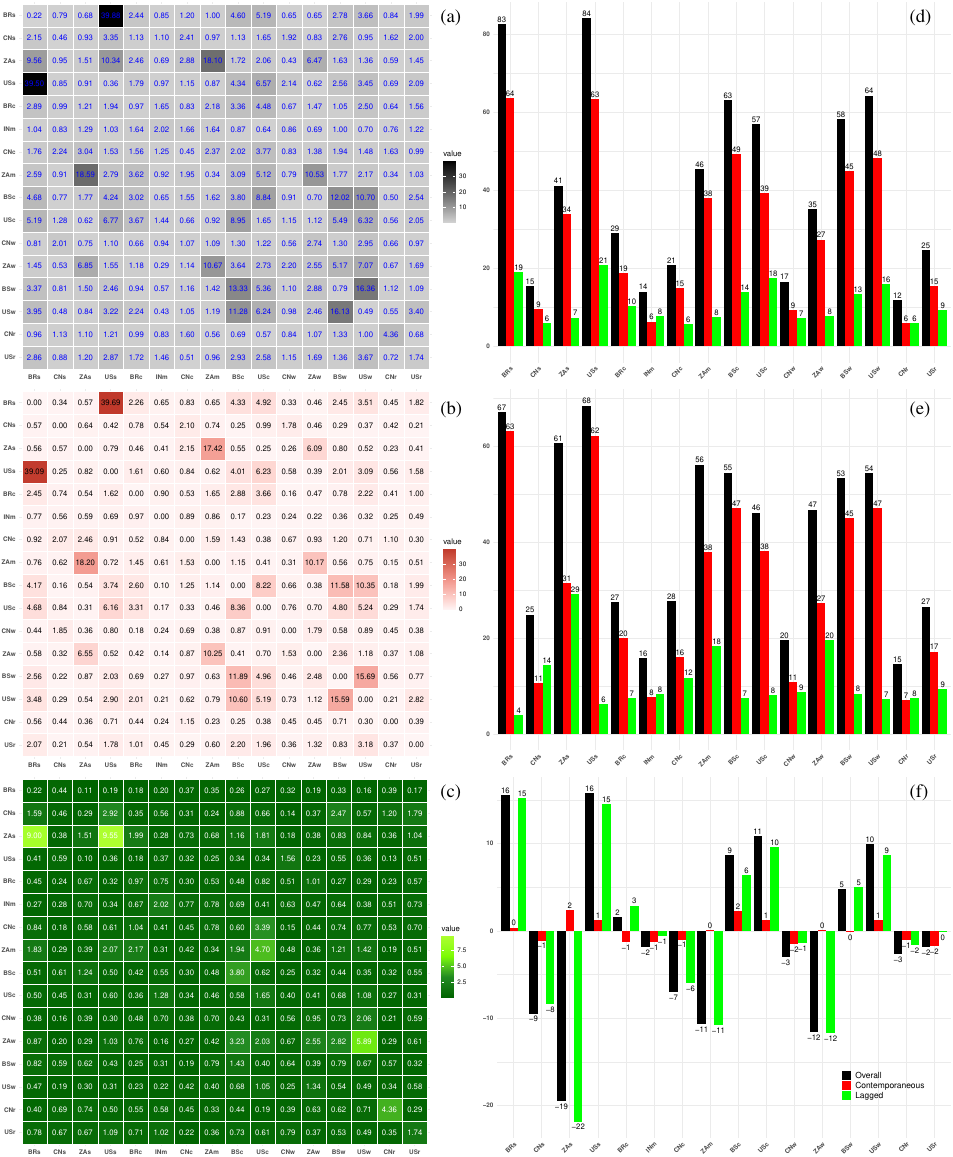}
    \caption{Averaged dynamic spillover indices of System 1. Notes: The $(i, j)$ element of the heatmap indicates the magnitude of the spillover effect from grain $j$ to grain $i$. Panels (a), (b), (c), (d), (e), and (f) represent the overall, contemporaneous, lagged, TO, FROM, and NET connectedness, respectively.}
    \label{Fig:Static:Connectedness}
\end{figure}

Our analysis starts with the average spillover effects for System 1, as depicted in Fig.~\ref{Fig:Static:Connectedness}. The average TCI stands at 41.54\%, with 30.55\% attributed to contemporaneous effects and 10.99\% to lagged dynamics.
From the perspectives of overall and contemporaneous connectedness, the spillover of risk between U.S. and Brazilian soybeans is the strongest, aligning with their highest positive correlation, as shown in Fig.~\ref{Fig:Correlation:US}. Lagged spillover effects are most pronounced between South African soybeans and U.S. and Brazilian soybeans. The TO and FROM indices for U.S. and Brazilian soybeans are the highest, while China’s rice exhibits the lowest.\footnote{Values before rounding for TO, FROM, and NET are provided in Table~\ref{Tb:Average:Dynamic:Connectedness:System1}.} 

Across overall, contemporaneous, and lagged spillover effects, Brazilian soybeans, U.S. soybeans, Black Sea corn, U.S. corn, Black Sea wheat, and U.S. wheat consistently function as net transmitters of shocks, while Chinese soybeans, Indian maize, Chinese corn, South African maize, Chinese wheat, South African wheat, and both Chinese and U.S. rice serve as net receivers \citep{Zhu-Dai-Zhou-2024-JFuturesMark}. Other grain futures, such as South African soybeans, show inconsistent roles as net transmitters or receivers across contemporaneous and lagged periods.

The overall TCI for System 2 is 23.56\%, as presented in Table~\ref{Tb:Average:Dynamic:Connectedness:System2}, with the strongest average spillover between South African soybeans and maize. South African staple grains and Brazilian soybeans contribute the most to spillovers to other markets, while South Africa’s grain market exhibits the strongest FROM spillover. In terms of overall NET connectedness, only Brazilian grains act as net transmitters of shocks.

\subsection{Time-varying total spillover effects}

\begin{figure}[htp]
	\centering
	\includegraphics[width=0.98\linewidth]{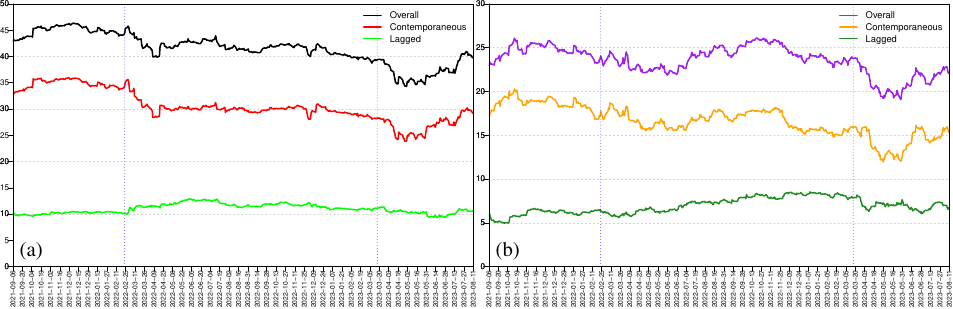}
	\caption{Dynamic total connectedness. (a) and (b) show the TCI of System 1 and System 2, with different colors indicating overall dynamic, contemporaneous, and lagged connectedness. The two vertical lines mark the Russia-Ukraine conflict's outbreak on February 24, 2022, and the start of the BSGI's extension period shortening from 120 days to 60 days on March 18, 2023.}
	\label{Fig:TCI}
\end{figure}

Fig.~\ref{Fig:TCI} illustrates the time-varying total connectedness, with the overall and contemporaneous TCI showing similar evolution patterns, and contemporaneous spillover effects markedly exceeding those in the lagged period.
The overall TCI of System 1 increased prior to the Russia-Ukraine conflict and began to decline a few days later, while the overall TCI of System 2 slightly dropped immediately after the conflict, followed by minor fluctuations. The TCI of both systems experienced a significant decline after the BSGI extension period was shortened.
Clearly, Figs.~\ref{Fig:Static:Connectedness} and \ref{Fig:TCI}, along with Table~\ref{Tb:Average:Dynamic:Connectedness:System2}, demonstrate that the overall, contemporaneous, and lagged connectedness of System 1 is significantly stronger than that of System 2, highlighting the substantial influence of the U.S. grain futures market on BRICS grain markets.

\subsection{Dynamic directional spillovers: FROM, TO, and NET}

\begin{figure}[H]
	\centering
	\includegraphics[width=0.73\linewidth]{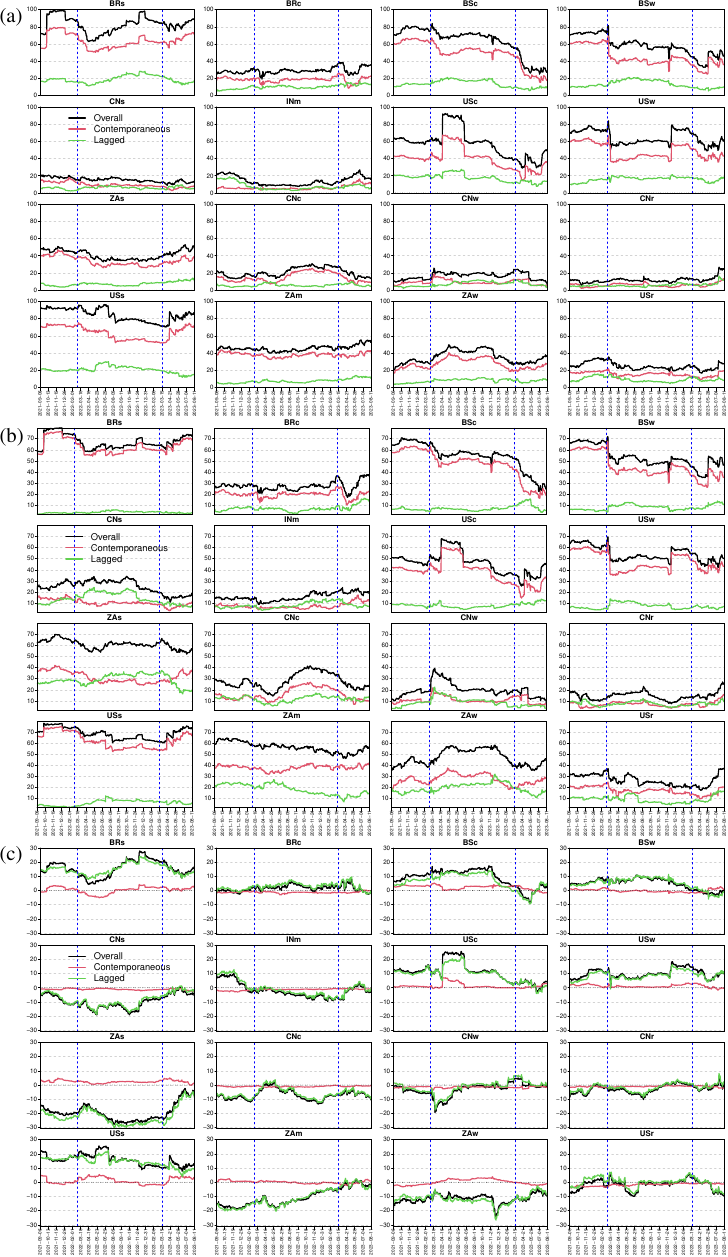}
	\caption{Directional connectedness of System 1. Panels (a), (b), and (c) represent the TO, FROM, and NET connectedness of System 1, respectively.}
	\label{Fig:TO:FROM:NET:US}
\end{figure}

Fig.~\ref{Fig:TO:FROM:NET:US} and Fig.~\ref{Fig:TO:FROM:NET:Only} depict the time-varying directional connectedness of System 1 and System 2, respectively. At a glance, most grains exhibit similar TO and FROM spillover trajectories, differing mainly in spillover intensity. The outbreak of the Russia-Ukraine conflict and the shortening of the BSGI extension period have inconsistent effects on the TO, FROM, and NET spillover effects across the 18 grain futures. The spillover from China's and India's grain futures to other markets is very low, generally below 20\%. Particularly, after the outbreak of the conflict, the spillover effects of U.S. and Black Sea wheat on other markets dropped by more than 20\%, with the spillover levels remaining stable for an extended period. Additionally, after the BSGI extension was shortened to 60 days, their spillovers also decreased by more than 20\%. U.S. corn saw its TO spillover reach 90\% for several months after the conflict broke out, before the BSGI came into effect. Compared to U.S. soybeans, corn, and wheat, the TO connectedness of U.S. rice is relatively weak. Therefore, even for different grain futures from the same country, the impact of the Russia-Ukraine conflict varies significantly. Correspondingly, the spillover from other markets received by Indian maize and Chinese rice is also very low, with the overall level remaining below 20\%.
For System 2, South African and Brazilian grain futures are the primary transmitters and receivers of risk spillovers, followed by Chinese corn. In contrast, Chinese rice has the lowest TO and FROM connectedness throughout the sample period, consistently below 10\%.

Regarding the time-varying NET spillover, it is observed that for both System 1 and System 2, their contemporaneous connectedness fluctuates around zero, indicating that each grain futures market experiences and contributes similar shocks in the same period. Naturally, the NET lagged connectedness follows a nearly identical trajectory to the overall connectedness.
Specifically, in System 1, Brazilian and U.S. soybeans, U.S. corn and wheat, followed by Brazilian corn, are consistently the highest net contributors of spillover. Before the shortening of the BSGI extension, Black Sea corn and wheat also remained net transmitters of shocks.
The South African soybeans, maize, wheat, and Chinese corn are consistently strong net receivers of shocks, while other grains, such as U.S. rice, fluctuate around zero, indicating extremely weak net spillovers.
For the net connectedness of intra-BRICS markets, only Brazilian soybeans and corn are consistent net transmitters of spillovers, while South African soybeans, maize, and Chinese soybeans are persistent net receivers of shocks. Among them, the net spillovers of Brazilian and South African soybeans far exceed those of other grains. The remaining grains, INm, ZAw, CNc, CNr, and CNw, exhibit weak net connectedness, fluctuating around zero.

\subsection{Net pairwise directional spillover effects}

\begin{figure}[htp]
			\centering
			\includegraphics[width=0.95\linewidth]{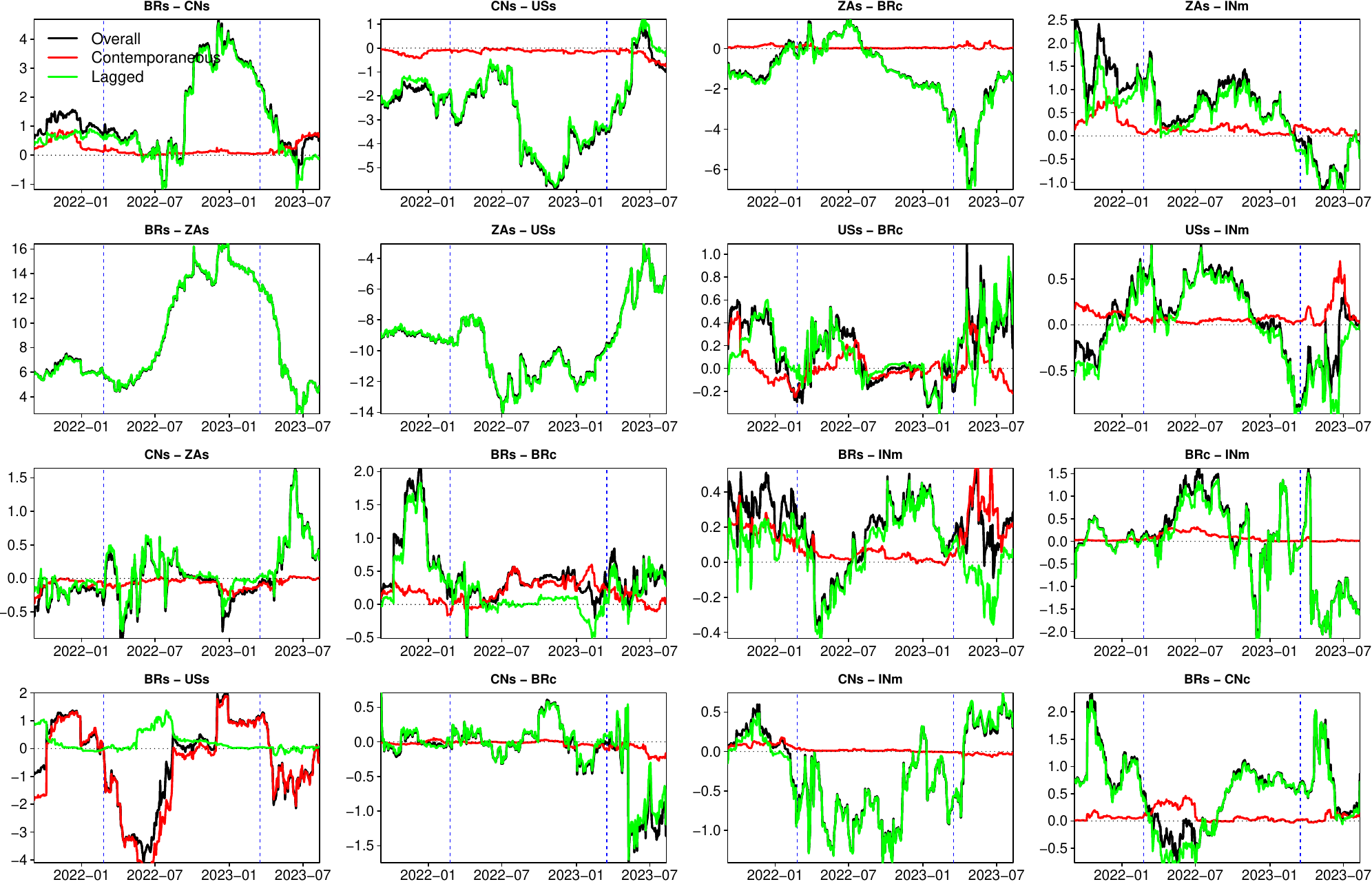}
			\caption{Net pairwise directional connectedness of System 1.}
			\label{Fig:NPDC:US}
\end{figure}

From the net pairwise directional spillover of System 1 presented in Fig.~\ref{Fig:NPDC:US}, it is evident that spillovers between soybean markets are the strongest. Brazil and the U.S. exhibit significant spillover effects to South Africa and China, with the spillover to South Africa reaching as high as 16\%. The NPDC between Brazilian and U.S. soybeans changes over time and is significantly influenced by the outbreak of the Russia-Ukraine conflict and the shortening of the BSGI. Additionally, the contemporaneous and lagged NPDC of each grain pair show substantial differences, with the contemporaneous NPDC of BRs and USs being stronger than the lagged one. Overall, the lagged NPDC dominates for most grain pairs. Additionally, most of the NPDC in System 2 follow similar patterns to those in System 1, but with varying spillover intensities. For instance, the NPDC between BRs and ZAs reaches a peak of 16\% in System 1, whereas it can reach up to 25\% in System 2.

\subsection{Impacts of Russia-Ukraine Conflict and BSGI}

Given the significant impact of the Russia-Ukraine conflict and the BSGI on the connectedness of grain futures, we divide the entire sample into three subsamples for a more in-depth analysis: Subsample 1, before the outbreak of the Russia-Ukraine conflict (before February 23, 2022); Subsample 2, from the onset of the conflict until the signing of the BSGI (from February 24, 2022, to July 21, 2022); and Subsample 3, after the BSGI came into effect (after July 22, 2022).

\begin{figure}[htp]
    \centering
    \includegraphics[width=0.95\linewidth]{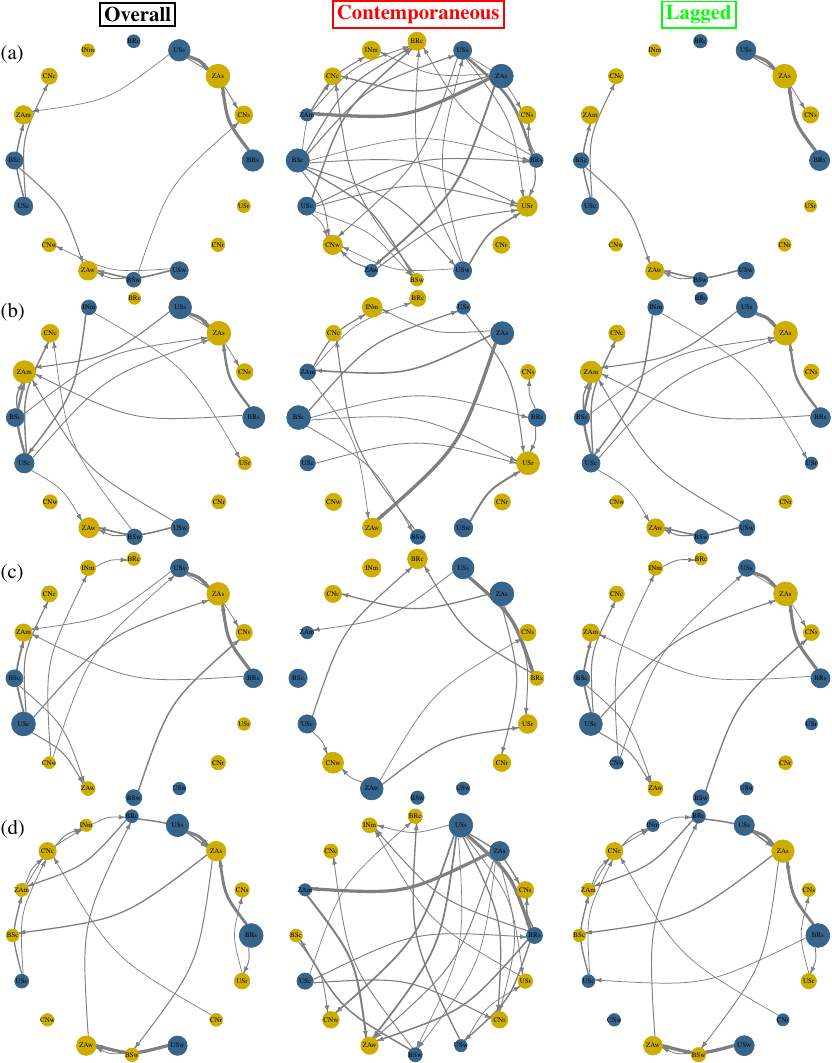}
    \caption{Net pairwise directional spillover networks of System 1. Notes: Edge thickness represents spillover strength. Blue and yellow nodes are transmitters and receivers, respectively. Panels (a) to (d) correspond to the full sample, Subsample 1, Subsample 2, and Subsample 3, showing overall, contemporaneous, and lagged dependency from left to right.}
    \label{Fig:Network:US}
\end{figure}

Fig.~\ref{Fig:Network:US} illustrates net pairwise directional spillover networks of System 1, with edges filtered using a threshold of 0.2 to highlight the major spillover effects. The lagged network at the whole sample level again confirms the strong net spillover from Brazilian and U.S. soybeans to South African soybeans. The contemporaneous network indicates that South African soybeans exhibit strong spillovers to wheat and maize, while U.S. soybeans also show significant spillovers to Brazilian soybeans. In contrast, the risk spillover contribution of Chinese rice to the entire system is very weak.

Affected by the Russia-Ukraine conflict and the BSGI, the spillover networks in the three subsamples show variations in transmitters, receivers, and spillover intensities. In the contemporaneous network of Subsample 2, BRs acts as a receiver, while in Subsamples 1 and 3, it is a transmitter. ZAw is a transmitter in Subsample 2 but a receiver in Subsamples 1 and 3. Similarly, in the lagged network of Subsample 2, BRc is a receiver, whereas it is a transmitter in Subsamples 1 and 3. In particular, after the signing of the BSGI, contemporaneous spillover interactions between grains in Subsample 3 are tightly connected, while those in Subsample 2 are sparse. Unlike in System 1, in System 2, the contemporaneous spillovers between grains in Subsample 2 are very widespread, while those in Subsamples 1 and 3 are quite sparse, clearly reflecting the impact of the Russia-Ukraine conflict on the grain market.

\subsection{Robustness test}	
	
\begin{figure}[H]
	\centering
	\includegraphics[width=0.98\linewidth]{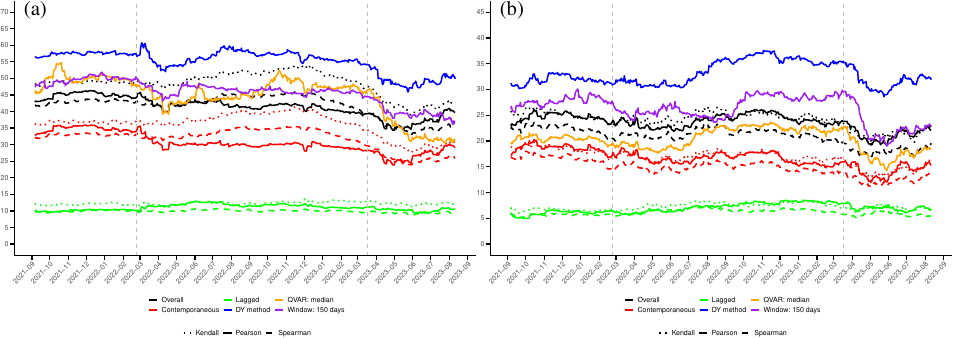}
	\caption{Robustness check for TCI. Panels (a) and (b) correspond to System 1 and System 2. Five methods are used to characterize TCI: $R^2$ decomposition with Kendall, Spearman, and Pearson correlations, and the DY and QVAR methods. A 150-day rolling window $R^2$ decomposition is also included for comparison.}
	\label{Fig:Robustness}
\end{figure}

Various robustness checks are conducted for both systems, as shown in Fig.~\ref{Fig:Robustness}, to evaluate the reliability of the findings. We use several approaches to assess the TCI: $R^2$ decomposition based on different correlations, the DY, and quantile VAR (QVAR). A 150-day rolling window is also included for comparison. The results from all six TCIs show nearly consistent evolution, confirming the robustness of the grain futures risk spillover results in each system.

\section{Conclusion}
\label{S4:conclusion}

Leveraging the advanced $R^2$ decomposition connectedness method, this study thoroughly investigates the contemporaneous and lagged spillover effects in the BRICS staple grain futures markets and their cross-market linkages with the U.S. markets. We find that contemporaneous spillovers dominate in both systems, while net spillovers are driven by lagged connectedness. Furthermore, the TCI reveals lower systemic risk in intra-BRICS markets compared to the system including the U.S., underscoring the U.S. grain market's significant influence on global grain futures.
System 1 indicates that Brazilian and U.S. soybeans, U.S. corn, wheat, and Brazilian corn are dominant net spillover contributors, with Black Sea corn and wheat also acting as net transmitters before the BSGI extension shortening. South African soybeans, maize, wheat, and Chinese corn are strong net receivers, while U.S. rice shows minimal spillovers near zero. For the net spillovers within intra-BRICS markets, the Brazilian market consistently acts as a transmitter of shocks, while South African soybeans and maize persistently serve as receivers of shocks. The net spillovers from China and India remain weak throughout the period. Additionally, we characterize the heterogeneous impacts of the Russia-Ukraine conflict and the BSGI on grain futures. Thus, practitioners can use spillover information from dominant net transmitters (e.g., soybean futures) to manage asset positions and adjust strategies based on investment horizons, particularly during crises or major policy announcements.

\section*{Acknowledgements}
	
This work was partly  supported by the National Social Science Foundation of China [Grant No. 24CGL072].

\newpage
\appendix
\section{Appendix}
\begin{landscape}
	\begin{table}[htbp] 
		\centering
		\caption{Averaged dynamic connectedness of System 1.}
		\label{Tb:Average:Dynamic:Connectedness:System1}
		{\fontsize{7.5pt}{9.5pt}\selectfont 
			\resizebox{1.5\textwidth}{!}{ 
				\begin{threeparttable}
					\begin{tabular}{llllllllllllllllll}
						\toprule
						& BRs & CNs & ZAs & USs & BRc & INm & CNc & ZAm & BSc & USc & CNw & ZAw & BSw & USw & CNr & USr & FROM\\
						\midrule
						BRs & 0.22 & 0.79 & 0.68 & 39.88 & 2.44 & 0.85 & 1.20 & 1.00 & 4.60 & 5.19 & 0.65 & 0.65 & 2.78 & 3.66 & 0.84 & 1.99 & 67.19\\
						& ( 0.00,  0.22) & ( 0.34,  0.44) & ( 0.57,  0.11) & (39.69,  0.19) & ( 2.26,  0.18) & ( 0.65,  0.20) & ( 0.83,  0.37) & ( 0.65,  0.35) & ( 4.33,  0.26) & ( 4.92,  0.27) & ( 0.33,  0.32) & ( 0.46,  0.19) & ( 2.45,  0.33) & ( 3.51,  0.16) & ( 0.45,  0.39) & ( 1.82,  0.17) & (63.26,  3.93)\\
						CNs & 2.15 & 0.46 & 0.93 & 3.35 & 1.13 & 1.10 & 2.41 & 0.97 & 1.13 & 1.65 & 1.92 & 0.83 & 2.76 & 0.95 & 1.62 & 2.00 & 24.90\\
						& ( 0.57,  1.59) & ( 0.00,  0.46) & ( 0.64,  0.29) & ( 0.42,  2.92) & ( 0.78,  0.35) & ( 0.54,  0.56) & ( 2.10,  0.31) & ( 0.74,  0.24) & ( 0.25,  0.88) & ( 0.99,  0.66) & ( 1.78,  0.14) & ( 0.46,  0.37) & ( 0.29,  2.47) & ( 0.37,  0.57) & ( 0.42,  1.20) & ( 0.21,  1.79) & (10.57, 14.33)\\
						ZAs & 9.56 & 0.95 & 1.51 & 10.34 & 2.46 & 0.69 & 2.88 & 18.10 & 1.72 & 2.06 & 0.43 & 6.47 & 1.63 & 1.36 & 0.59 & 1.45 & 60.70\\
						& ( 0.56,  9.00) & ( 0.57,  0.38) & ( 0.00,  1.51) & ( 0.79,  9.55) & ( 0.46,  1.99) & ( 0.41,  0.28) & ( 2.15,  0.73) & (17.42,  0.68) & ( 0.55,  1.16) & ( 0.25,  1.81) & ( 0.26,  0.18) & ( 6.09,  0.38) & ( 0.80,  0.83) & ( 0.52,  0.84) & ( 0.23,  0.36) & ( 0.41,  1.04) & (31.47, 29.23)\\
						USs & 39.50 & 0.85 & 0.91 & 0.36 & 1.79 & 0.97 & 1.15 & 0.87 & 4.34 & 6.57 & 2.14 & 0.62 & 2.56 & 3.45 & 0.69 & 2.09 & 68.50\\
						& (39.09,  0.41) & ( 0.25,  0.59) & ( 0.82,  0.10) & ( 0.00,  0.36) & ( 1.61,  0.18) & ( 0.60,  0.37) & ( 0.84,  0.32) & ( 0.62,  0.25) & ( 4.01,  0.34) & ( 6.23,  0.34) & ( 0.58,  1.56) & ( 0.39,  0.23) & ( 2.01,  0.55) & ( 3.09,  0.36) & ( 0.56,  0.13) & ( 1.58,  0.51) & (62.26,  6.24)\\
						BRc & 2.89 & 0.99 & 1.21 & 1.94 & 0.97 & 1.65 & 0.83 & 2.18 & 3.36 & 4.48 & 0.67 & 1.47 & 1.05 & 2.50 & 0.64 & 1.56 & 27.43\\
						& ( 2.45,  0.45) & ( 0.74,  0.24) & ( 0.54,  0.67) & ( 1.62,  0.32) & ( 0.00,  0.97) & ( 0.90,  0.75) & ( 0.53,  0.30) & ( 1.65,  0.53) & ( 2.88,  0.48) & ( 3.66,  0.82) & ( 0.16,  0.51) & ( 0.47,  1.01) & ( 0.78,  0.27) & ( 2.22,  0.29) & ( 0.41,  0.23) & ( 1.00,  0.57) & (20.00,  7.43)\\
						INm & 1.04 & 0.83 & 1.29 & 1.03 & 1.64 & 2.02 & 1.66 & 1.64 & 0.87 & 0.64 & 0.86 & 0.69 & 1.00 & 0.70 & 0.76 & 1.22 & 15.87\\
						& ( 0.77,  0.27) & ( 0.56,  0.28) & ( 0.59,  0.70) & ( 0.69,  0.34) & ( 0.97,  0.67) & ( 0.00,  2.02) & ( 0.89,  0.77) & ( 0.86,  0.78) & ( 0.17,  0.69) & ( 0.23,  0.41) & ( 0.24,  0.63) & ( 0.22,  0.47) & ( 0.36,  0.64) & ( 0.32,  0.38) & ( 0.25,  0.51) & ( 0.49,  0.73) & ( 7.62,  8.26)\\
						CNc & 1.76 & 2.24 & 3.04 & 1.53 & 1.56 & 1.25 & 0.45 & 2.37 & 2.02 & 3.77 & 0.83 & 1.38 & 1.94 & 1.48 & 1.63 & 0.99 & 27.78\\
						& ( 0.92,  0.84) & ( 2.07,  0.18) & ( 2.46,  0.58) & ( 0.91,  0.61) & ( 0.52,  1.04) & ( 0.84,  0.41) & ( 0.00,  0.45) & ( 1.59,  0.78) & ( 1.43,  0.60) & ( 0.38,  3.39) & ( 0.67,  0.15) & ( 0.93,  0.44) & ( 1.20,  0.74) & ( 0.71,  0.77) & ( 1.10,  0.53) & ( 0.30,  0.70) & (16.03, 11.76)\\
						ZAm & 2.59 & 0.91 & 18.59 & 2.79 & 3.62 & 0.92 & 1.95 & 0.34 & 3.09 & 5.12 & 0.79 & 10.53 & 1.77 & 2.17 & 0.34 & 1.03 & 56.20\\
						& ( 0.76,  1.83) & ( 0.62,  0.29) & (18.20,  0.39) & ( 0.72,  2.07) & ( 1.45,  2.17) & ( 0.61,  0.31) & ( 1.53,  0.42) & ( 0.00,  0.34) & ( 1.15,  1.94) & ( 0.41,  4.70) & ( 0.31,  0.48) & (10.17,  0.36) & ( 0.56,  1.21) & ( 0.75,  1.42) & ( 0.15,  0.19) & ( 0.51,  0.51) & (37.90, 18.30)\\
						BSc & 4.68 & 0.77 & 1.77 & 4.24 & 3.02 & 0.65 & 1.55 & 1.62 & 3.80 & 8.84 & 0.91 & 0.70 & 12.02 & 10.70 & 0.50 & 2.54 & 54.51\\
						& ( 4.17,  0.51) & ( 0.16,  0.61) & ( 0.54,  1.24) & ( 3.74,  0.50) & ( 2.60,  0.42) & ( 0.10,  0.55) & ( 1.25,  0.30) & ( 1.14,  0.48) & ( 0.00,  3.80) & ( 8.22,  0.62) & ( 0.66,  0.25) & ( 0.38,  0.32) & (11.58,  0.44) & (10.35,  0.35) & ( 0.18,  0.32) & ( 1.99,  0.55) & (47.06,  7.45)\\
						USc & 5.19 & 1.28 & 0.62 & 6.77 & 3.67 & 1.44 & 0.66 & 0.92 & 8.95 & 1.65 & 1.15 & 1.12 & 5.49 & 6.32 & 0.56 & 2.05 & 46.19\\
						& ( 4.68,  0.50) & ( 0.84,  0.45) & ( 0.31,  0.31) & ( 6.16,  0.60) & ( 3.31,  0.36) & ( 0.17,  1.28) & ( 0.33,  0.34) & ( 0.46,  0.46) & ( 8.36,  0.58) & ( 0.00,  1.65) & ( 0.76,  0.40) & ( 0.70,  0.41) & ( 4.80,  0.68) & ( 5.24,  1.08) & ( 0.29,  0.27) & ( 1.74,  0.31) & (38.16,  8.03)\\
						CNw & 0.81 & 2.01 & 0.75 & 1.10 & 0.66 & 0.94 & 1.07 & 1.09 & 1.30 & 1.22 & 0.56 & 2.74 & 1.30 & 2.95 & 0.66 & 0.97 & 19.57\\
						& ( 0.44,  0.38) & ( 1.85,  0.16) & ( 0.36,  0.39) & ( 0.80,  0.30) & ( 0.18,  0.48) & ( 0.24,  0.70) & ( 0.69,  0.38) & ( 0.38,  0.70) & ( 0.87,  0.43) & ( 0.91,  0.31) & ( 0.00,  0.56) & ( 1.79,  0.95) & ( 0.58,  0.73) & ( 0.89,  2.06) & ( 0.45,  0.21) & ( 0.38,  0.59) & (10.80,  8.77)\\
						ZAw & 1.45 & 0.53 & 6.85 & 1.55 & 1.18 & 0.29 & 1.14 & 10.67 & 3.64 & 2.73 & 2.20 & 2.55 & 5.17 & 7.07 & 0.67 & 1.69 & 46.82\\
						& ( 0.58,  0.87) & ( 0.32,  0.20) & ( 6.55,  0.29) & ( 0.52,  1.03) & ( 0.42,  0.76) & ( 0.14,  0.16) & ( 0.87,  0.27) & (10.25,  0.42) & ( 0.41,  3.23) & ( 0.70,  2.03) & ( 1.53,  0.67) & ( 0.00,  2.55) & ( 2.36,  2.82) & ( 1.18,  5.89) & ( 0.37,  0.29) & ( 1.08,  0.61) & (27.28, 19.54)\\
						BSw & 3.37 & 0.81 & 1.50 & 2.46 & 0.94 & 0.57 & 1.16 & 1.42 & 13.33 & 5.36 & 1.10 & 2.88 & 0.79 & 16.36 & 1.12 & 1.09 & 53.46\\
						& ( 2.56,  0.82) & ( 0.22,  0.59) & ( 0.87,  0.62) & ( 2.03,  0.43) & ( 0.69,  0.25) & ( 0.27,  0.31) & ( 0.97,  0.19) & ( 0.63,  0.79) & (11.89,  1.43) & ( 4.96,  0.40) & ( 0.46,  0.64) & ( 2.48,  0.39) & ( 0.00,  0.79) & (15.69,  0.67) & ( 0.56,  0.57) & ( 0.77,  0.32) & (45.04,  8.43)\\
						USw & 3.95 & 0.48 & 0.84 & 3.22 & 2.24 & 0.43 & 1.05 & 1.19 & 11.28 & 6.24 & 0.98 & 2.46 & 16.13 & 0.49 & 0.55 & 3.40 & 54.42\\
						& ( 3.48,  0.47) & ( 0.29,  0.19) & ( 0.54,  0.30) & ( 2.90,  0.31) & ( 2.01,  0.23) & ( 0.21,  0.22) & ( 0.62,  0.42) & ( 0.79,  0.40) & (10.60,  0.68) & ( 5.19,  1.05) & ( 0.73,  0.25) & ( 1.12,  1.34) & (15.59,  0.54) & ( 0.00,  0.49) & ( 0.21,  0.34) & ( 2.82,  0.58) & (47.10,  7.32)\\
						CNr & 0.96 & 1.13 & 1.10 & 1.21 & 0.99 & 0.83 & 1.60 & 0.56 & 0.69 & 0.57 & 0.84 & 1.07 & 1.33 & 1.00 & 4.36 & 0.68 & 14.55\\
						& ( 0.56,  0.40) & ( 0.44,  0.69) & ( 0.36,  0.74) & ( 0.71,  0.50) & ( 0.44,  0.55) & ( 0.24,  0.58) & ( 1.15,  0.45) & ( 0.23,  0.33) & ( 0.25,  0.44) & ( 0.38,  0.19) & ( 0.45,  0.39) & ( 0.45,  0.63) & ( 0.71,  0.62) & ( 0.30,  0.71) & ( 0.00,  4.36) & ( 0.39,  0.29) & ( 7.04,  7.51)\\
						USr & 2.86 & 0.88 & 1.20 & 2.87 & 1.72 & 1.46 & 0.51 & 0.96 & 2.93 & 2.58 & 1.15 & 1.69 & 1.36 & 3.67 & 0.72 & 1.74 & 26.56\\
						& ( 2.07,  0.78) & ( 0.21,  0.67) & ( 0.54,  0.67) & ( 1.78,  1.09) & ( 1.01,  0.71) & ( 0.45,  1.02) & ( 0.29,  0.22) & ( 0.60,  0.36) & ( 2.20,  0.73) & ( 1.96,  0.61) & ( 0.36,  0.79) & ( 1.32,  0.37) & ( 0.83,  0.53) & ( 3.18,  0.49) & ( 0.37,  0.35) & ( 0.00,  1.74) & (17.18,  9.38)\\\\
						\hline
						TO & 82.76 & 15.44 & 41.27 & 84.27 & 29.06 & 14.05 & 20.82 & 45.55 & 63.24 & 57.01 & 16.62 & 35.29 & 58.29 & 64.36 & 11.87 & 24.75 & 664.66\\
						& ( 63.66,  19.11) & (  9.47,   5.97) & ( 33.87,   7.40) & ( 63.49,  20.77) & ( 18.73,  10.33) & (  6.35,   7.70) & ( 15.04,   5.78) & ( 38.00,   7.55) & ( 49.35,  13.89) & ( 39.41,  17.61) & (  9.27,   7.35) & ( 27.42,   7.87) & ( 44.89,  13.40) & ( 48.32,  16.04) & (  6.00,   5.88) & ( 15.49,   9.27) & (488.75, 175.91)\\
						Inc.Own & 82.98 & 15.90 & 42.78 & 84.62 & 30.03 & 16.07 & 21.27 & 45.89 & 67.04 & 58.67 & 17.18 & 37.84 & 59.08 & 64.85 & 16.23 & 26.50 & TCI\\
						& (63.66, 19.32) & ( 9.47,  6.43) & (33.87,  8.91) & (63.49, 21.13) & (18.73, 11.30) & ( 6.35,  9.72) & (15.04,  6.23) & (38.00,  7.89) & (49.35, 17.69) & (39.41, 19.26) & ( 9.27,  7.91) & (27.42, 10.42) & (44.89, 14.19) & (48.32, 16.53) & ( 6.00, 10.23) & (15.49, 11.01) & ($\rm TCI^C$, $\rm TCI^L$)\\
						NET & 15.57 & -9.46 & -19.43 & 15.77 & 1.63 & -1.83 & -6.96 & -10.65 & 8.73 & 10.82 & -2.95 & -11.53 & 4.83 & 9.94 & -2.68 & -1.81 & 41.54\\
						& ( 0.39,  15.18) & (-1.10,  -8.36) & ( 2.39, -21.82) & ( 1.23,  14.54) & (-1.27,   2.90) & (-1.27,  -0.56) & (-0.99,  -5.97) & ( 0.10, -10.75) & ( 2.29,   6.44) & ( 1.25,   9.57) & (-1.53,  -1.42) & ( 0.14, -11.67) & (-0.15,   4.98) & ( 1.22,   8.72) & (-1.04,  -1.64) & (-1.69,  -0.12) & (30.55, 10.99)\\
						\bottomrule
					\end{tabular}
		\end{threeparttable}}}
	\end{table}
\end{landscape}

\begin{landscape}
	\begin{table}[htbp] 
		\centering
		\caption{Averaged dynamic connectedness of intra-BRICS markets.}
		\label{Tb:Average:Dynamic:Connectedness:System2}
		{\fontsize{7.6pt}{10pt}\selectfont 
			\begin{threeparttable}
				\begin{tabular}{lllllllllllll}
					\toprule
					& BRs & CNs & ZAs & BRc & INm & CNc & ZAm & CNw & ZAw & CNr & FROM\\
					\midrule
					BRs & 0.46 & 0.98 & 1.32 & 6.71 & 1.47 & 2.39 & 1.46 & 1.21 & 0.85 & 1.43 & 17.83\\
					& ( 0.00,  0.46) & ( 0.54,  0.44) & ( 1.14,  0.18) & ( 6.41,  0.30) & ( 1.20,  0.27) & ( 1.69,  0.70) & ( 0.82,  0.64) & ( 0.62,  0.59) & ( 0.45,  0.40) & ( 0.83,  0.60) & (13.72,  4.11)\\
					CNs & 3.78 & 0.47 & 1.19 & 1.06 & 1.03 & 2.43 & 1.19 & 1.84 & 0.46 & 1.60 & 14.58\\
					& ( 0.56,  3.22) & ( 0.00,  0.47) & ( 0.96,  0.23) & ( 0.72,  0.34) & ( 0.59,  0.44) & ( 2.07,  0.35) & ( 0.99,  0.20) & ( 1.69,  0.15) & ( 0.26,  0.20) & ( 0.58,  1.02) & ( 8.42,  6.16)\\
					ZAs & 16.15 & 1.25 & 1.78 & 2.72 & 0.77 & 3.31 & 21.08 & 0.42 & 7.11 & 0.60 & 53.42\\
					& ( 0.97, 15.19) & ( 0.88,  0.38) & ( 0.00,  1.78) & ( 0.49,  2.23) & ( 0.44,  0.33) & ( 2.42,  0.90) & (20.46,  0.62) & ( 0.28,  0.14) & ( 6.71,  0.40) & ( 0.21,  0.39) & (32.85, 20.57)\\
					BRc & 6.85 & 0.97 & 1.32 & 0.89 & 1.75 & 1.11 & 2.41 & 0.68 & 1.71 & 0.74 & 17.53\\
					& ( 6.44,  0.40) & ( 0.70,  0.27) & ( 0.55,  0.78) & ( 0.00,  0.89) & ( 0.90,  0.85) & ( 0.78,  0.32) & ( 1.93,  0.47) & ( 0.17,  0.51) & ( 0.69,  1.02) & ( 0.46,  0.28) & (12.62,  4.90)\\
					INm & 1.54 & 0.88 & 1.33 & 1.58 & 2.13 & 1.64 & 1.64 & 0.81 & 0.69 & 0.77 & 10.88\\
					& ( 1.22,  0.32) & ( 0.60,  0.28) & ( 0.59,  0.75) & ( 0.94,  0.65) & ( 0.00,  2.13) & ( 0.89,  0.75) & ( 0.78,  0.85) & ( 0.23,  0.58) & ( 0.19,  0.50) & ( 0.22,  0.55) & ( 5.65,  5.23)\\
					CNc & 2.84 & 2.21 & 3.37 & 2.19 & 1.30 & 0.49 & 2.95 & 0.79 & 1.65 & 1.58 & 18.88\\
					& ( 1.66,  1.17) & ( 2.06,  0.15) & ( 2.77,  0.60) & ( 0.78,  1.41) & ( 0.86,  0.44) & ( 0.00,  0.49) & ( 2.19,  0.77) & ( 0.68,  0.11) & ( 1.28,  0.38) & ( 1.04,  0.53) & (13.32,  5.56)\\
					ZAm & 4.11 & 1.12 & 21.59 & 4.63 & 0.89 & 2.47 & 0.28 & 0.76 & 14.27 & 0.33 & 50.17\\
					& ( 0.70,  3.41) & ( 0.86,  0.26) & (21.14,  0.45) & ( 1.69,  2.94) & ( 0.59,  0.30) & ( 2.03,  0.44) & ( 0.00,  0.28) & ( 0.28,  0.48) & (13.74,  0.53) & ( 0.14,  0.19) & (41.18,  8.99)\\
					CNw & 0.93 & 1.88 & 0.81 & 0.78 & 1.00 & 1.08 & 1.05 & 0.47 & 3.19 & 0.66 & 11.40\\
					& ( 0.64,  0.29) & ( 1.72,  0.16) & ( 0.37,  0.43) & ( 0.17,  0.61) & ( 0.23,  0.77) & ( 0.70,  0.39) & ( 0.33,  0.72) & ( 0.00,  0.47) & ( 2.26,  0.94) & ( 0.48,  0.18) & ( 6.90,  4.50)\\
					ZAw & 1.79 & 0.42 & 7.58 & 1.56 & 0.34 & 1.70 & 15.04 & 2.48 & 3.42 & 0.74 & 31.64\\
					& ( 0.37,  1.41) & ( 0.22,  0.20) & ( 7.25,  0.34) & ( 0.64,  0.92) & ( 0.14,  0.21) & ( 1.25,  0.45) & (14.61,  0.43) & ( 1.99,  0.48) & ( 0.00,  3.42) & ( 0.46,  0.28) & (26.93,  4.72)\\
					CNr & 1.10 & 1.23 & 0.96 & 1.14 & 0.75 & 1.51 & 0.51 & 0.85 & 1.23 & 4.26 & 9.27\\
					& ( 0.84,  0.25) & ( 0.59,  0.64) & ( 0.29,  0.67) & ( 0.47,  0.66) & ( 0.22,  0.54) & ( 1.08,  0.43) & ( 0.20,  0.31) & ( 0.47,  0.37) & ( 0.52,  0.71) & ( 0.00,  4.26) & ( 4.69,  4.59)\\\\
					\hline
					TO & 39.09 & 10.94 & 39.47 & 22.37 & 9.32 & 17.65 & 47.33 & 9.84 & 31.15 & 8.45 & 235.61\\
					& ( 13.42, 25.67) & (  8.17,  2.78) & ( 35.06,  4.41) & ( 12.30, 10.06) & (  5.17,  4.15) & ( 12.91,  4.74) & ( 42.31,  5.02) & (  6.43,  3.41) & ( 26.09,  5.06) & (  4.42,  4.03) & (166.28, 69.33)\\
					Inc.Own & 39.55 & 11.41 & 41.25 & 23.25 & 11.45 & 18.14 & 47.60 & 10.30 & 34.57 & 12.71 & TCI\\
					& (13.42, 26.13) & ( 8.17,  3.25) & (35.06,  6.19) & (12.30, 10.95) & ( 5.17,  6.28) & (12.91,  5.23) & (42.31,  5.29) & ( 6.43,  3.88) & (26.09,  8.48) & ( 4.42,  8.29) & ($\rm TCI^C$, $\rm TCI^L$)\\
					NET & 21.26 & -3.64 & -13.95 & 4.84 & -1.56 & -1.23 & -2.85 & -1.56 & -0.49 & -0.82 & 23.56\\
					& (-0.30,  21.56) & (-0.25,  -3.38) & ( 2.21, -16.16) & (-0.32,   5.16) & (-0.48,  -1.08) & (-0.41,  -0.82) & ( 1.12,  -3.97) & (-0.47,  -1.09) & (-0.84,   0.34) & (-0.26,  -0.56) & (16.63, 6.93)\\
					\bottomrule
				\end{tabular}
				\begin{tablenotes}[para,flushleft]
					\scriptsize Notes: $R^{2}$ decomposed measures are based on a 200-day rolling-window VAR model with a lag length of order one (BIC). Values in parentheses represent contemporaneous and lagged effects, respectively.
				\end{tablenotes}
		\end{threeparttable}}
	\end{table}
\end{landscape}

\begin{figure}[h]
	\centering
	\includegraphics[width=0.65\linewidth]{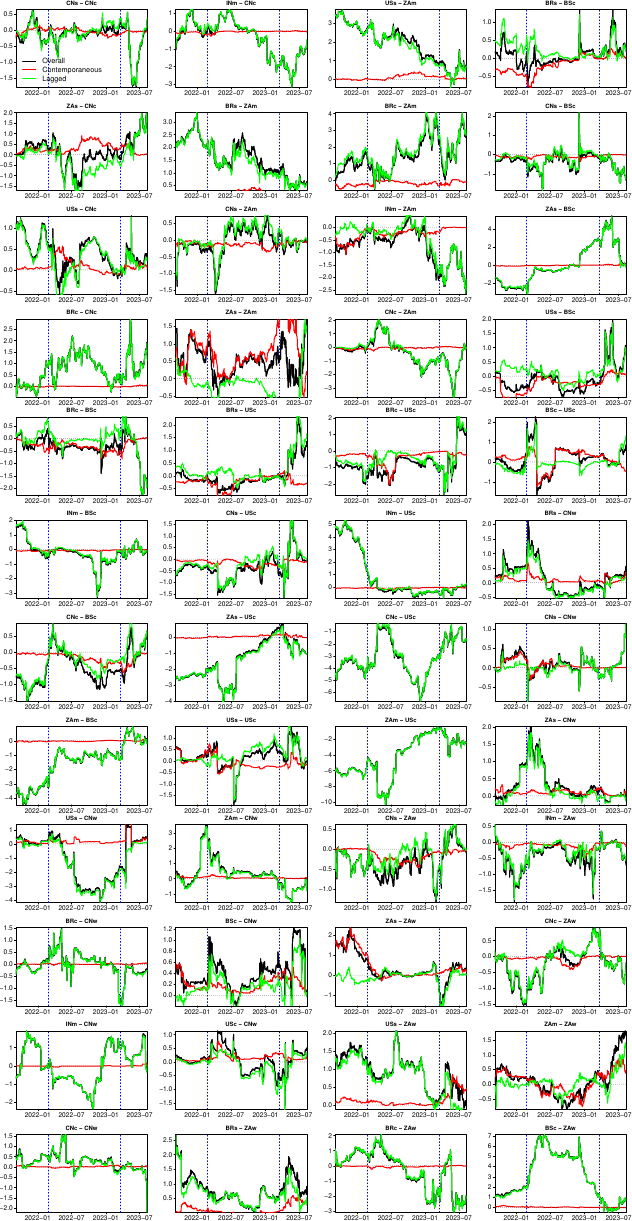}
	\caption{Continue...}
\end{figure}
\begin{figure}[p]\ContinuedFloat
	\centering
	\includegraphics[width=0.65\linewidth]{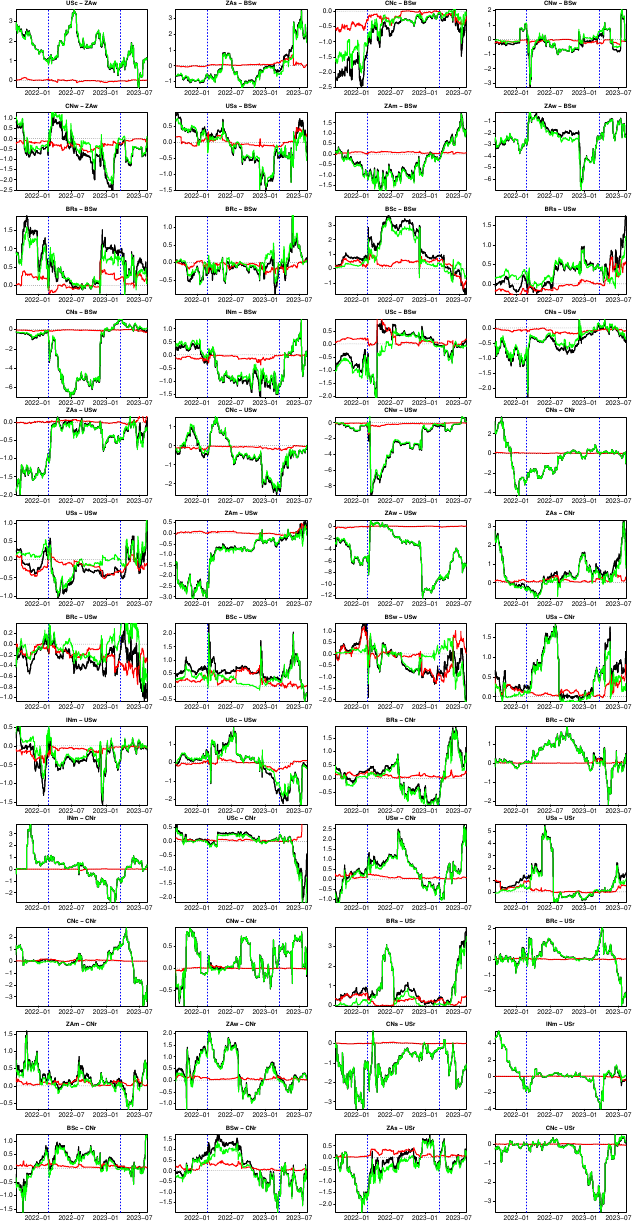}
	\caption{Continue...}
\end{figure}

\begin{figure}[t]\ContinuedFloat
	\centering
	\includegraphics[width=0.7\linewidth]{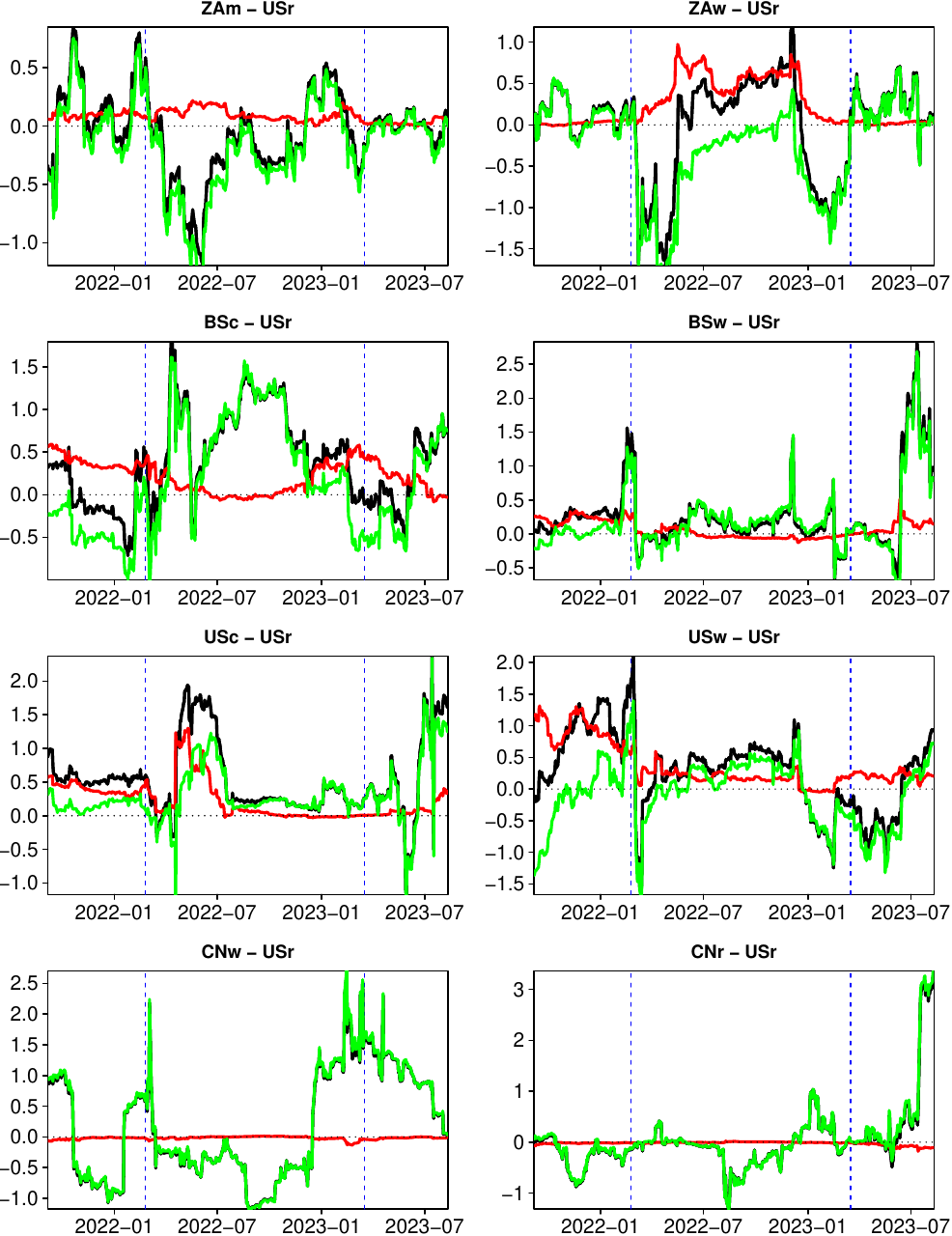}
	\caption{Net pairwise directional connectedness of System 1.}
	\label{Fig:NET:System1:Rest}
\end{figure}

\begin{figure}[h]
	\centering
	\includegraphics[width=0.58\linewidth]{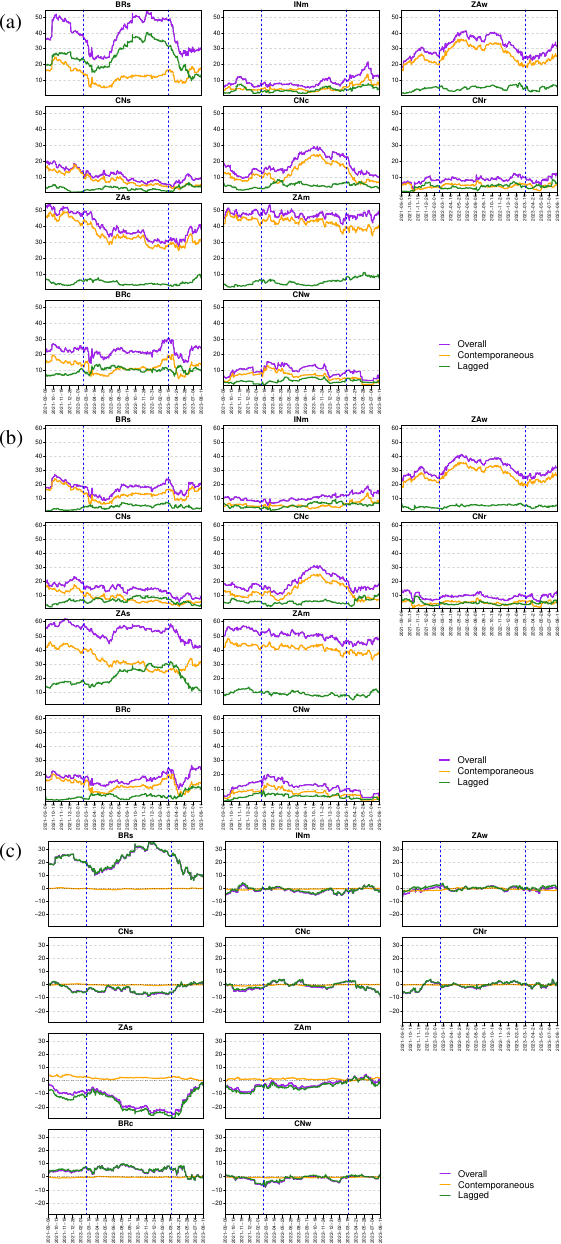}
	\caption{Directional connectedness. Panels (a), (b), and (c) represent the TO, FROM, and NET connectedness of intra-BRICS markets (System 2), respectively.}
	\label{Fig:TO:FROM:NET:Only}
\end{figure}

\begin{figure}[h]
	\centering
	\includegraphics[width=0.66\linewidth]{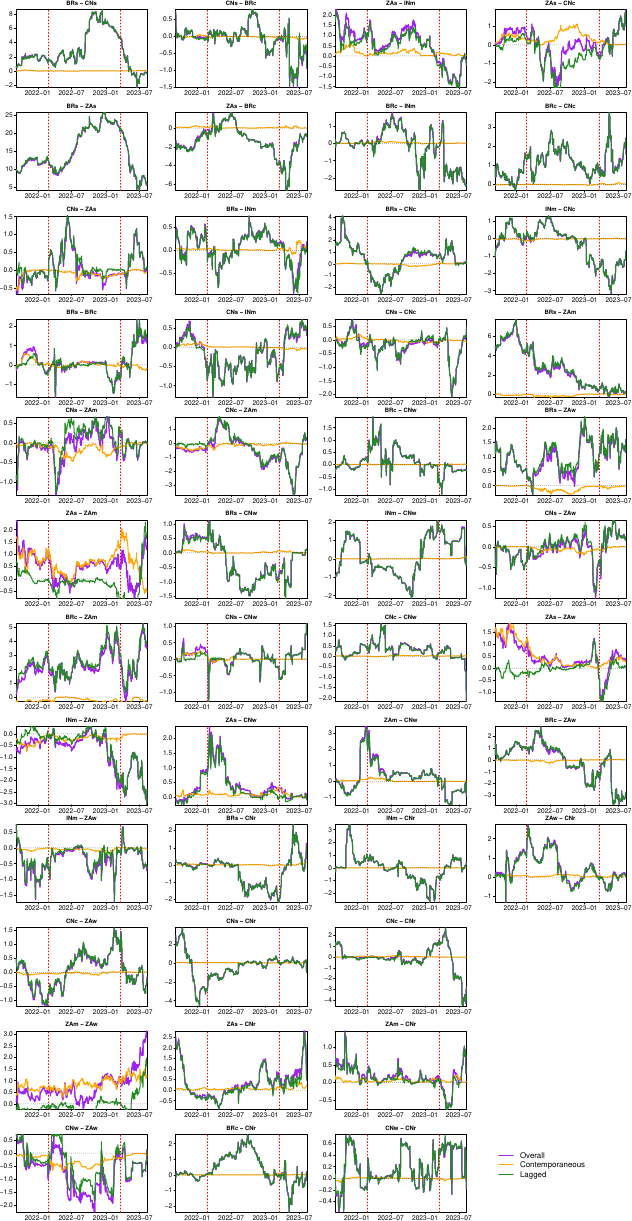}
	\caption{Net pairwise directional connectedness of System 2.}
	\label{Fig:NPDC:Only}
\end{figure}

\begin{figure}[h]
	\centering
	\includegraphics[width=0.95\linewidth]{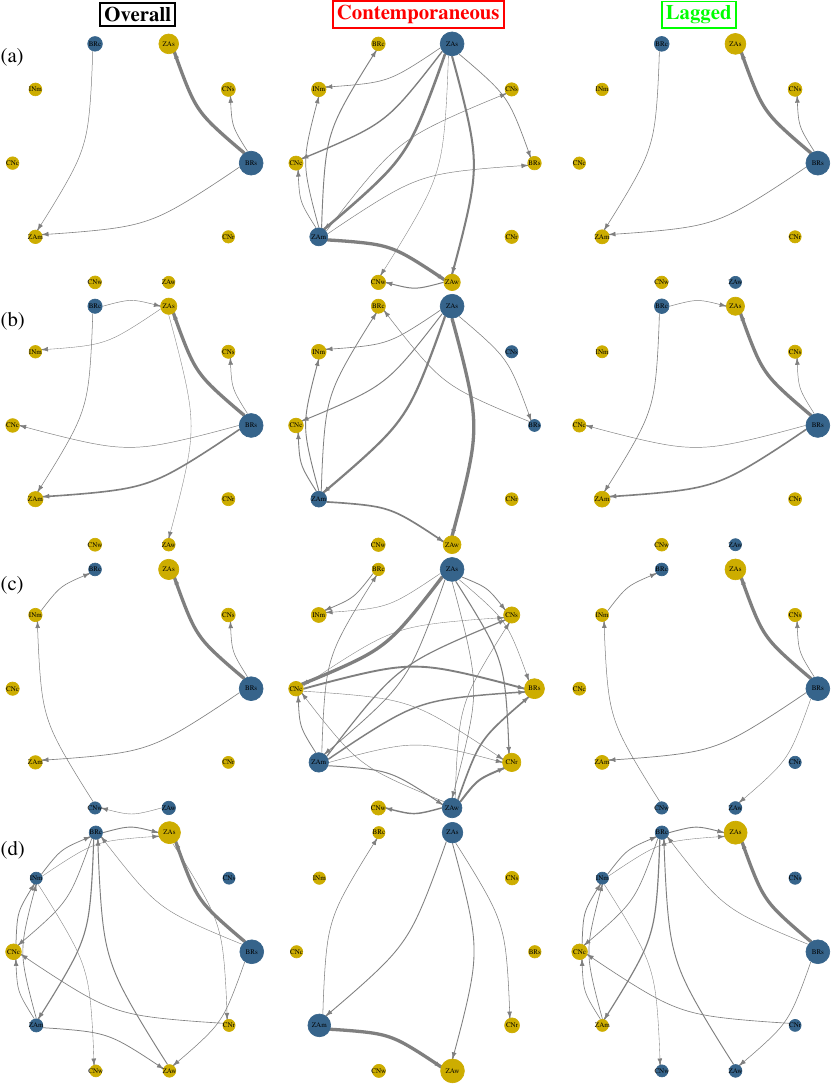}
	\caption{Net pairwise directional spillover networks of System 2. Notes: Edge thickness represents spillover strength. Blue and yellow nodes are transmitters and receivers, respectively. Panels (a) to (d) correspond to the full sample, Subsample 1, Subsample 2, and Subsample 3, showing overall, contemporaneous, and lagged dependency from left to right.}
	\label{Fig:Network:Only}
\end{figure} 


\begin{thebibliography}{22}
	\expandafter\ifx\csname natexlab\endcsname\relax\def\natexlab#1{#1}\fi
	\providecommand{\url}[1]{\texttt{#1}}
	\providecommand{\href}[2]{#2}
	\providecommand{\path}[1]{#1}
	\providecommand{\DOIprefix}{doi:}
	\providecommand{\ArXivprefix}{arXiv:}
	\providecommand{\URLprefix}{URL: }
	\providecommand{\Pubmedprefix}{pmid:}
	\providecommand{\doi}[1]{\href{http://dx.doi.org/#1}{\path{#1}}}
	\providecommand{\Pubmed}[1]{\href{pmid:#1}{\path{#1}}}
	\providecommand{\bibinfo}[2]{#2}
	\ifx\xfnm\relax \def\xfnm[#1]{\unskip,\space#1}\fi
	\bibitem[{Balli et~al.(2023)Balli, Balli, Dang and
		Gabauer}]{Balli-Balli-Dang-Gabauer-2023-FinancResLett}
	\bibinfo{author}{Balli, F.}, \bibinfo{author}{Balli, H.O.},
	\bibinfo{author}{Dang, T.H.N.}, \bibinfo{author}{Gabauer, D.},
	\bibinfo{year}{2023}.
	\newblock \bibinfo{title}{Contemporaneous and lagged {$R^2$} decomposed
		connectedness approach: New evidence from the energy futures market}.
	\newblock \bibinfo{journal}{Financ. Res. Lett.} \bibinfo{volume}{57},
	\bibinfo{pages}{104168}.
	\newblock \DOIprefix\doi{10.1016/j.frl.2023.104168}.
	\bibitem[{Behnassi and El~Haiba(2022)}]{Behnassi-ElHaiba-2022-NatHumBehav}
	\bibinfo{author}{Behnassi, M.}, \bibinfo{author}{El~Haiba, M.},
	\bibinfo{year}{2022}.
	\newblock \bibinfo{title}{Implications of the {R}ussia-{U}kraine war for global
		food security}.
	\newblock \bibinfo{journal}{Nat. Hum. Behav.} \bibinfo{volume}{6},
	\bibinfo{pages}{754--755}.
	\newblock \DOIprefix\doi{10.1038/s41562-022-01391-x}.
	\bibitem[{Chen and Weng(2018)}]{Chen-Weng-2018-EmergMarkFinancTrade}
	\bibinfo{author}{Chen, Q.}, \bibinfo{author}{Weng, X.}, \bibinfo{year}{2018}.
	\newblock \bibinfo{title}{Information flows between the {US} and {C}hina’s
		agricultural commodity futures markets—based on {VAR-BEKK-S}kew-t model}.
	\newblock \bibinfo{journal}{Emerg. Mark. Financ. Trade} \bibinfo{volume}{54},
	\bibinfo{pages}{71--87}.
	\newblock \DOIprefix\doi{10.1080/1540496X.2016.1230492}.
	\bibitem[{Cocca et~al.(2024)Cocca, Gabauer and
		Pomberger}]{Cocca-Gabauer-Pomberger-2024-EnergyEcon}
	\bibinfo{author}{Cocca, T.}, \bibinfo{author}{Gabauer, D.},
	\bibinfo{author}{Pomberger, S.}, \bibinfo{year}{2024}.
	\newblock \bibinfo{title}{Clean energy market connectedness and investment
		strategies: New evidence from {DCC}-{GARCH} {$R^2$} decomposed connectedness
		measures}.
	\newblock \bibinfo{journal}{Energy Econ.} \bibinfo{volume}{136},
	\bibinfo{pages}{107680}.
	\newblock \DOIprefix\doi{10.1016/j.eneco.2023.107680}.
	\bibitem[{Goldstein and Yang(2022)}]{Goldstein-Yang-2022-JFinanc}
	\bibinfo{author}{Goldstein, I.}, \bibinfo{author}{Yang, L.},
	\bibinfo{year}{2022}.
	\newblock \bibinfo{title}{Commodity financialization and information
		transmission}.
	\newblock \bibinfo{journal}{J. Financ.} \bibinfo{volume}{77},
	\bibinfo{pages}{2613--2667}.
	\newblock \DOIprefix\doi{10.1111/jofi.13165}.
	\bibitem[{Guo et~al.(2023)Guo, Li, Zhang, Ji and
		Zhao}]{Guo-Li-Zhang-Ji-Zhao-2023-JCommodMark}
	\bibinfo{author}{Guo, K.}, \bibinfo{author}{Li, Y.}, \bibinfo{author}{Zhang,
		Y.}, \bibinfo{author}{Ji, Q.}, \bibinfo{author}{Zhao, W.},
	\bibinfo{year}{2023}.
	\newblock \bibinfo{title}{How are climate risk shocks connected to agricultural
		markets?}
	\newblock \bibinfo{journal}{J. Commod. Mark.} \bibinfo{volume}{32},
	\bibinfo{pages}{100367}.
	\newblock \DOIprefix\doi{10.1016/j.jcomm.2023.100367}.
	\bibitem[{Han et~al.(2013)Han, Liang and
		Tang}]{Han-Liang-Tang-2013-QuantFinanc}
	\bibinfo{author}{Han, L.}, \bibinfo{author}{Liang, R.}, \bibinfo{author}{Tang,
		K.}, \bibinfo{year}{2013}.
	\newblock \bibinfo{title}{{Cross-market soybean futures price discovery: does
			the Dalian Commodity Exchange affect the Chicago Board of Trade?}}
	\newblock \bibinfo{journal}{Quant. Financ.} \bibinfo{volume}{13},
	\bibinfo{pages}{613--626}.
	\newblock \DOIprefix\doi{10.1080/14697688.2013.775477}.
	\bibitem[{Hu et~al.(2024)Hu, Zhu, Zhang and
		Zeng}]{Hu-Zhu-Zhang-Zeng-2024-PlosOne}
	\bibinfo{author}{Hu, X.}, \bibinfo{author}{Zhu, B.}, \bibinfo{author}{Zhang,
		B.}, \bibinfo{author}{Zeng, L.}, \bibinfo{year}{2024}.
	\newblock \bibinfo{title}{Extreme risk spillovers between {US} and {C}hinese
		agricultural futures markets in crises: A dependence-switching copula-{CoVaR}
		model}.
	\newblock \bibinfo{journal}{PLoS One} \bibinfo{volume}{19},
	\bibinfo{pages}{e0299237}.
	\newblock \DOIprefix\doi{10.1371/journal.pone.0299237}.
	\bibitem[{Jiang et~al.(2017)Jiang, Todorova, Roca and
		Su}]{Jiang-Todorova-Roca-Su-2017-ApplEcon}
	\bibinfo{author}{Jiang, H.}, \bibinfo{author}{Todorova, N.},
	\bibinfo{author}{Roca, E.}, \bibinfo{author}{Su, J.J.}, \bibinfo{year}{2017}.
	\newblock \bibinfo{title}{Dynamics of volatility transmission between the
		{U}.{S}. and the {C}hinese agricultural futures markets}.
	\newblock \bibinfo{journal}{Appl. Econ.} \bibinfo{volume}{49},
	\bibinfo{pages}{3435--3452}.
	\newblock \DOIprefix\doi{10.1080/00036846.2016.1262517}.
	\bibitem[{Just and Echaust(2022)}]{Just-Echaust-2022-EconLett}
	\bibinfo{author}{Just, M.}, \bibinfo{author}{Echaust, K.},
	\bibinfo{year}{2022}.
	\newblock \bibinfo{title}{Dynamic spillover transmission in agricultural
		commodity markets: {W}hat has changed after the {COVID}-19 threat?}
	\newblock \bibinfo{journal}{Econ. Lett.} \bibinfo{volume}{217},
	\bibinfo{pages}{110671}.
	\newblock \DOIprefix\doi{10.1016/j.econlet.2022.110671}.
	\bibitem[{Khalfaoui et~al.(2023)Khalfaoui, Hammoudeh and
		Rehman}]{Khalfaoui-Hammoudeh-Rehman-2023-EmergMarkRev}
	\bibinfo{author}{Khalfaoui, R.}, \bibinfo{author}{Hammoudeh, S.},
	\bibinfo{author}{Rehman, M.Z.}, \bibinfo{year}{2023}.
	\newblock \bibinfo{title}{Spillovers and connectedness among {BRICS} stock
		markets, cryptocurrencies, and uncertainty: Evidence from the quantile vector
		autoregression network}.
	\newblock \bibinfo{journal}{Emerg. Mark. Rev.} \bibinfo{volume}{54},
	\bibinfo{pages}{101002}.
	\newblock \DOIprefix\doi{10.1016/j.ememar.2023.101002}.
	\bibitem[{Li and Hayes(2017)}]{Li-Hayes-2017-JFuturesMark}
	\bibinfo{author}{Li, C.}, \bibinfo{author}{Hayes, D.J.}, \bibinfo{year}{2017}.
	\newblock \bibinfo{title}{Price discovery on the international soybean futures
		markets: A threshold co-integration approach}.
	\newblock \bibinfo{journal}{J. Futures Mark.} \bibinfo{volume}{37},
	\bibinfo{pages}{52--70}.
	\newblock \DOIprefix\doi{10.1002/fut.21794}.
	\bibitem[{Li and Chavas(2023)}]{Li-Chavas-2023-AmJAgrEcon}
	\bibinfo{author}{Li, J.}, \bibinfo{author}{Chavas, J.P.}, \bibinfo{year}{2023}.
	\newblock \bibinfo{title}{A dynamic analysis of the distribution of commodity
		futures and spot prices}.
	\newblock \bibinfo{journal}{Am. J. Agr. Econ.} \bibinfo{volume}{105},
	\bibinfo{pages}{122--143}.
	\newblock \DOIprefix\doi{10.1111/ajae.12309}.
	\bibitem[{Naeem et~al.(2024)Naeem, Chatziantoniou, Gabauer and
		Karim}]{Naeem-Chatziantoniou-Gabauer-Karim-2024-IntRevFinancAnal}
	\bibinfo{author}{Naeem, M.A.}, \bibinfo{author}{Chatziantoniou, I.},
	\bibinfo{author}{Gabauer, D.}, \bibinfo{author}{Karim, S.},
	\bibinfo{year}{2024}.
	\newblock \bibinfo{title}{Measuring the {G}20 stock market return transmission
		mechanism: Evidence from the {$R^2$} connectedness approach}.
	\newblock \bibinfo{journal}{Int. Rev. Financ. Anal.} \bibinfo{volume}{91},
	\bibinfo{pages}{102986}.
	\newblock \DOIprefix\doi{10.1016/j.irfa.2023.102986}.
	\bibitem[{Ni et~al.(2024)Ni, Cherif and
		Chen}]{Ni-Cherif-Chen-2024-FinancResLett}
	\bibinfo{author}{Ni, G.}, \bibinfo{author}{Cherif, H.H.},
	\bibinfo{author}{Chen, Z.}, \bibinfo{year}{2024}.
	\newblock \bibinfo{title}{Measuring dynamic spillovers between crude oil and
		grain commodity markets: A comparative analysis of demand and supply shocks}.
	\newblock \bibinfo{journal}{Financ. Res. Lett.} \bibinfo{volume}{67},
	\bibinfo{pages}{105748}.
	\newblock \DOIprefix\doi{10.1016/j.frl.2024.105748}.
	\bibitem[{Urak et~al.(2024)Urak, Bilgic, Florkowski and
		Bozma}]{Urak-Bilgic-Florkowski-Bozma-2024-BorsaIstanbRev}
	\bibinfo{author}{Urak, F.}, \bibinfo{author}{Bilgic, A.},
	\bibinfo{author}{Florkowski, W.J.}, \bibinfo{author}{Bozma, G.},
	\bibinfo{year}{2024}.
	\newblock \bibinfo{title}{Confluence of {COVID}-19 and the {R}ussia-{U}kraine
		conflict: Effects on agricultural commodity prices and food security}.
	\newblock \bibinfo{journal}{Borsa Istanb. Rev.} \bibinfo{volume}{24},
	\bibinfo{pages}{506--519}.
	\newblock \DOIprefix\doi{10.1016/j.bir.2024.02.008}.
	\bibitem[{Wang et~al.(2024)Wang, Dong, Sun, Shi and
		Ji}]{Wang-Dong-Sun-Shi-Ji-2024-EconModel}
	\bibinfo{author}{Wang, H.}, \bibinfo{author}{Dong, Y.}, \bibinfo{author}{Sun,
		M.}, \bibinfo{author}{Shi, B.}, \bibinfo{author}{Ji, H.},
	\bibinfo{year}{2024}.
	\newblock \bibinfo{title}{Dynamic dependence of futures basis between the
		{C}hinese and international grains markets}.
	\newblock \bibinfo{journal}{Econ. Model.} \bibinfo{volume}{130},
	\bibinfo{pages}{106584}.
	\newblock \DOIprefix\doi{10.1016/j.econmod.2023.106584}.
	\bibitem[{Wu et~al.(2023)Wu, Ren, Wan and
		Liu}]{Wu-Ren-Wan-Liu-2023-FinancResLett}
	\bibinfo{author}{Wu, Y.}, \bibinfo{author}{Ren, W.}, \bibinfo{author}{Wan, J.},
	\bibinfo{author}{Liu, X.}, \bibinfo{year}{2023}.
	\newblock \bibinfo{title}{Time-frequency volatility connectedness between
		fossil energy and agricultural commodities: Comparing the {COVID}-19 pandemic
		with the {R}ussia-{U}kraine conflict}.
	\newblock \bibinfo{journal}{Financ. Res. Lett.} \bibinfo{volume}{55},
	\bibinfo{pages}{103866}.
	\newblock \DOIprefix\doi{10.1016/j.frl.2023.103866}.
	\bibitem[{Zhang et~al.(2024)Zhang, Sun, Shi, Ding and
		Zhao}]{Zhang-Sun-Shi-Ding-Zhao-2024-HumSocSciCommun}
	\bibinfo{author}{Zhang, Y.}, \bibinfo{author}{Sun, Y.}, \bibinfo{author}{Shi,
		H.}, \bibinfo{author}{Ding, S.}, \bibinfo{author}{Zhao, Y.},
	\bibinfo{year}{2024}.
	\newblock \bibinfo{title}{{COVID}-19, the {R}ussia-{U}kraine war and the
		connectedness between the {US} and {C}hinese agricultural futures markets}.
	\newblock \bibinfo{journal}{Hum. Soc. Sci. Commun.} \bibinfo{volume}{11},
	\bibinfo{pages}{1--15}.
	\newblock \DOIprefix\doi{10.1057/s41599-024-02852-6}.
	\bibitem[{Zhou et~al.(2024)Zhou, Dai, Duong and
		Dai}]{Zhou-Dai-Duong-Dai-2024-JEconBehavOrgan}
	\bibinfo{author}{Zhou, W.X.}, \bibinfo{author}{Dai, Y.S.},
	\bibinfo{author}{Duong, K.T.}, \bibinfo{author}{Dai, P.F.},
	\bibinfo{year}{2024}.
	\newblock \bibinfo{title}{The impact of the {R}ussia-{U}kraine conflict on the
		extreme risk spillovers between agricultural futures and spots}.
	\newblock \bibinfo{journal}{J. Econ. Behav. Organ.} \bibinfo{volume}{217},
	\bibinfo{pages}{91--111}.
	\newblock \DOIprefix\doi{10.1016/j.jebo.2023.11.004}.
	\bibitem[{Zhu et~al.(2024)Zhu, Dai and Zhou}]{Zhu-Dai-Zhou-2024-JFuturesMark}
	\bibinfo{author}{Zhu, H.Y.}, \bibinfo{author}{Dai, P.F.},
	\bibinfo{author}{Zhou, W.X.}, \bibinfo{year}{2024}.
	\newblock \bibinfo{title}{Uncovering the {S}ino-{US} dynamic risk spillovers
		effects: Evidence from agricultural futures markets}.
	\newblock \bibinfo{journal}{J. Futures Mark.} \bibinfo{volume}{44},
	\bibinfo{pages}{1888--1910}.
	\newblock \DOIprefix\doi{10.1002/fut.22551}.
	\bibitem[{{\v{Z}}ivkov et~al.(2020){\v{Z}}ivkov, Kuzman and
		Subi{\'c}}]{Zivkov-Kuzman-Subic-2020-AgricEcon}
	\bibinfo{author}{{\v{Z}}ivkov, D.}, \bibinfo{author}{Kuzman, B.},
	\bibinfo{author}{Subi{\'c}, J.}, \bibinfo{year}{2020}.
	\newblock \bibinfo{title}{What bayesian quantiles can tell about volatility
		transmission between the major agricultural futures?}
	\newblock \bibinfo{journal}{Agric. Econ.} \bibinfo{volume}{66},
	\bibinfo{pages}{215--225}.
	\newblock \DOIprefix\doi{10.17221/127/2019-AGRICECON}.
	
\end{thebibliography}
\end{document}